# Spin-dependent Tunneling Barriers in CoPc/VSe$_2$ from Many-Body Interactions

*Runrun Xu[1,†], Fengyuan Xuan[2,†], and Su Ying Quek[1,2]\**

[1]Department of Physics, National University of Singapore, 2 Science Drive 3, 117542, Singapore

[2]Centre for Advanced 2D Materials, National University of Singapore, 6 Science Drive 2, 117546, Singapore

\* To whom correspondence should be addressed: phyqsy@nus.edu.sg

†: These authors contributed equally to this work.

## Abstract

Mixed-dimensional magnetic heterostructures are intriguing, newly available platforms to explore quantum physics and its applications. Using state-of-the-art many-body perturbation theory, we predict the energy level alignment for a self-assembled monolayer of cobalt phthalocyanine (CoPc) molecules on magnetic $VSe_2$ monolayers. The predicted projected density of states on CoPc agrees with experimental scanning tunneling spectra. Consistent with experiment, we predict a shoulder in the unoccupied region of the spectra that is absent from mean-field calculations. Unlike the nearly spin-degenerate gas phase frontier molecular orbitals, the tunneling barriers at the interface are spin-dependent, a finding of interest for quantum information and spintronics applications. Both the experimentally observed shoulder and the predicted spin-dependent tunneling barriers originate from many-body interactions in the interface-hybridized states. Our results showcase the intricate many-body physics that governs the properties of these mixed-dimensional magnetic heterostructures, and suggests the possibility of manipulating the spin-dependent tunneling barriers through modifications of interface coupling.

**TOC Graphic**

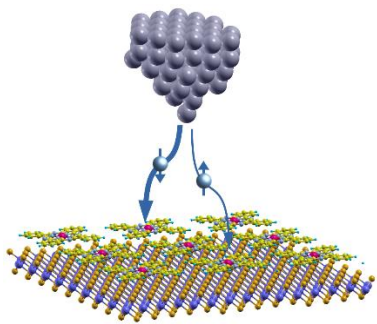

**Main Text**

Magnetic materials and their interfaces with other systems are fundamental to the phenomena of giant magnetoresistance and tunneling magnetoresistance,[1] which have revolutionized the data storage industry. On the other hand, magnetic molecules on substrates are emerging as a potential scalable alternative to solid state spin qubits for quantum computing.[2] Retaining the magnetic moment on supported magnetic molecules can be challenging - for example, the spin on cobalt phthalocyanine molecules (CoPc) is quenched when CoPc molecules are adsorbed on Pb(111)[3-4] and Au(111).[5] The use of two-dimensional (2D) layered substrates without dangling bonds can potentially reduce the interaction at the interface and minimize such spin quenching effects.

With rapid advancements in the synthesis and characterization of mixed dimensional heterostructures,[6] it is timely to study mixed-dimensional *magnetic* heterostructures to investigate their potential in spintronics and quantum computing. The tunneling barriers of spin-polarized electrons in such systems not only governs tunneling magnetoresistance, but are also critical components of quantum computing architectures based on zero-dimensional quantum molecular systems.[7] For example, spin-to-charge conversion used to read out spin quantum states relies on a difference in tunneling barriers for the two spin states.[8] Given the importance of spin-dependent tunneling barriers and lifetimes in such applications, it is critical to obtain a complete understanding of the spin splitting and energy level alignment (ELA) at the interfaces of these systems. First principles predictions of these quantities are particularly useful, because experimental measurements such as spin-polarized scanning tunneling spectroscopy are difficult to come by. Furthermore, *ab initio* approaches enable a detailed understanding of structure-property relationships at a level that is difficult to parallel in experiment.

Unfortunately, density functional theory (DFT) with standard local and semi-local exchange-correlation functionals fails to predict an ELA close to experimental values. The gas-phase gap between the highest occupied molecular orbital (HOMO) and the lowest unoccupied molecular orbital (LUMO) is typically underestimated by several eVs using standard DFT. Self-interaction corrected DFT[9] and hybrid functionals can improve this value, but does not account for non-local screening effects from the substrate/environment, which can reduce the HOMO-LUMO gap significantly.[10-14] Screening effects from a metal substrate have recently been captured using DFT calculations with an optimally tuned range-separated hybrid functional,[15] but an *ab initio* approach to determine this functional for more complex heterostructures is lacking. In contrast, many-body perturbation theory within the GW approximation[16] provides a rigorous framework to predict the ELA at interfaces. However, GW calculations are computationally demanding and are typically not used to explore large magnetic interfaces. Numerous efforts have been directed at predicting the ELA without the full computational expense of a GW calculation.[10-11, 13, 17-21] One draw-back of these approaches is that they are designed for physisorbed systems and do not take into account effects of interface hybridization.

We have recently developed a GW approach (called XAF-GW)[12] that enables tractable GW calculations of large interface systems, including the effects of interface hybridization.[12] In this work, we adapt the XAF-GW method to enable the prediction of spin-dependent ELA at complex interfaces, and apply the approach to study the spin splitting and ELA for a self-assembled monolayer of CoPc molecules on a 2D $VSe_2$ magnetic[22-23] monolayer. Transition metal phthalocyanine (TMPc) molecules are molecules of choice in molecular spintronics[24-27] because of their inherent stability, tunability, scalability and magnetic properties. Our predicted projected density of states (PDOS) agrees with the scanning tunneling spectroscopy (STS)[28] data for the same system. Consistent with experiment, we predict a shoulder in the unoccupied region of the spectra that is absent from DFT results. In contrast to the nearly spin-degenerate gas phase frontier molecular orbitals, the tunneling barriers at the interface are spin-dependent. We show that both the experimentally observed shoulder and the predicted spin-dependent tunneling barriers originate from many-body self-energy effects on the interface-hybridized states, underlining the intricate many-body physics at work in these mixed dimensional systems. The magnetic moments on CoPc and $VSe_2$ are both slightly enhanced in the heterostructure, and the spins on Co and V exhibit antiparallel alignment. Our work shows that TMPcs on 2D layered magnetic substrates are promising candidates for molecular spintronics and quantum computing applications, and suggests that the spin-dependent tunneling barriers can be controlled by tuning the interface hybridization, paving the way for bottom-up design of functional magnetic heterostructures.

High resolution scanning tunnelling microscopy images have shown that CoPc molecules self-assemble to form an ordered lattice when deposited on a $VSe_2$ monolayer that is supported on graphite.[28] Based on the arrangement and orientation of molecules in this lattice,[28] we choose a supercell for CoPc/$VSe_2$ as shown in Figure 1a. This choice of lattice minimizes the strain on both the experimentally derived CoPc lattice and the $VSe_2$ substrate, and is in reasonable agreement with experiment. Three possible adsorption sites for the CoPc molecules are considered (Figure 1b). After geometry optimization (see SI for details of methods), the adsorption energies are -3.58 eV, -3.68 eV and -3.86 eV for the V, Se and H sites, respectively. Figure 1f shows the relaxed atomic structure of the most stable configuration (H site), which will be used for this study.

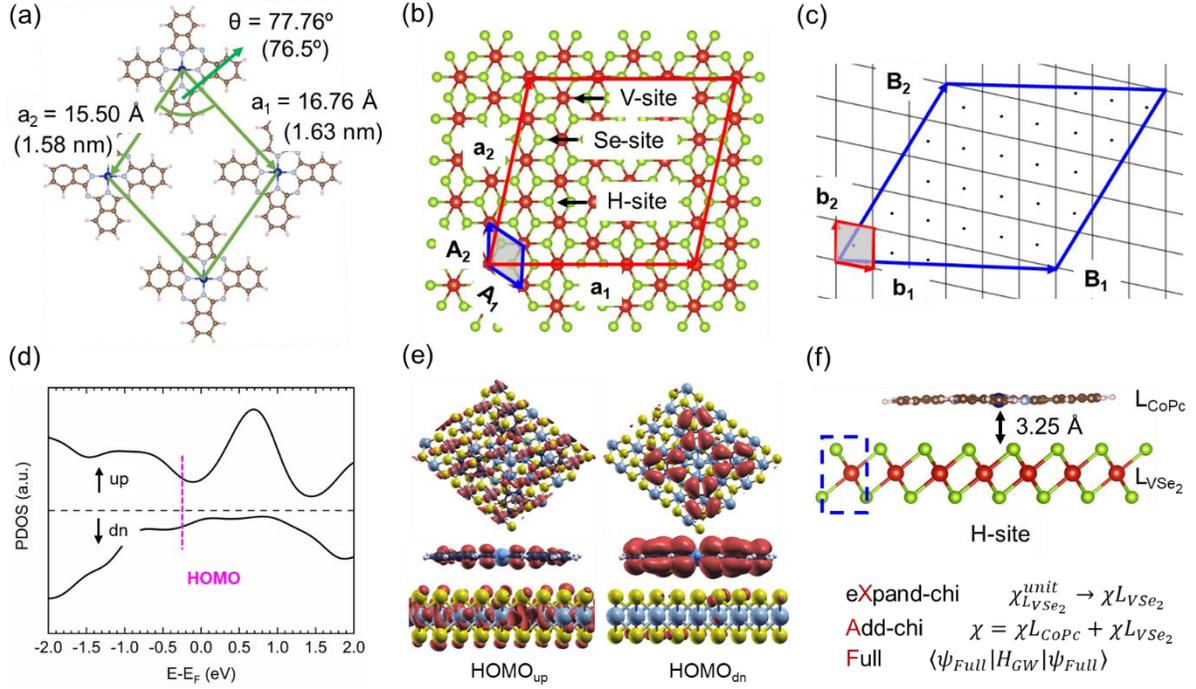

Figure 1. (a) Arrangement of CoPc molecules in self-assembled monolayer on VSe$_2$. Lattice parameters from experiment[28] (in brackets) match closely with the lattice parameters used in the calculations. (b) Real space VSe$_2$ lattice. The supercell for the CoPc/VSe$_2$ heterostructure (red) and the unit cell (u.c.; blue) used for the eXpand-chi step in XAF-GW (see Methods), are indicated. V, Se and H sites refer to the lateral positions of Co for different adsorption sites considered here. (c) Reciprocal lattice unit cells corresponding to the CoPc/VSe$_2$ supercell (red) and to the real space u.c. of VSe$_2$ (blue) in (b). (d) DFT+U projected density of states of VSe$_2$ in CoPc/VSe$_2$. The DFT+U HOMO ($a_{1u}$) energy in CoPc/VSe$_2$ is marked with a dashed line. (e) CoPc/VSe$_2$ heterostructure wavefunctions corresponding to the HOMO level marked in (d). HOMO$_{up}$: spin up HOMO, HOMO$_{dn}$: spin down HOMO. The isosurface value is 1% of the maximum value. (f) Schematic figure explaining the XAF-GW approach in CoPc/VSe$_2$. The relaxed atomic geometry for the most stable adsorption site (H-site) is shown.

Unlike CoPc on Pb(111)[3-4] and Au(111),[5] the magnetic moment on adsorbed CoPc is not quenched, but is in fact slightly enhanced compared to the gas phase (by ~ 0.03 $\mu_B$/f.u). The magnetic moment on VSe$_2$ is also enhanced by the same amount. This information suggests that 2D layered magnetic substrates are ideal for preserving and even enhancing the spins on single molecule magnets. Furthermore, it is found that the spins on Co and V favour an antiparallel alignment, as shown in Fig. 2a. This can be understood from the fact that the spin on Se in VSe$_2$ is antiparallel to that in V due to through-bond kinetic exchange[29-30] while the spins on the surface Se atoms are ferromagnetically coupled to those in Co because of Coulomb exchange.[30] This Coulomb exchange is consistent with the Pauli pushback effect[31] observed at the interface of CoPc and VSe$_2$, where electrons from the VSe$_2$ surface are pushed back by Pauli repulsive interaction with electrons from CoPc (Fig. 2b).

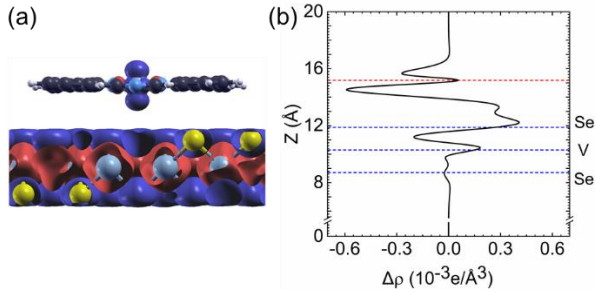

Figure 2. (a) Spin density, and (b) planar-averaged charge density redistribution at the interface of CoPc/VSe$_2$. The spin density is defined as $\rho_{spin} = \rho_{\uparrow,CoPc/VSe_2} - \rho_{\downarrow,CoPc/VSe_2}$. Spin up: red, spin down: blue. Antiparallel spin alignment between V and Co can be observed. The isosurface value is 1% of the maximum value. The charge density redistribution is defined as $\Delta\rho = \rho_{CoPc/VSe_2} - \rho_{CoPc} - \rho_{VSe_2}$ (where $\rho = \rho_\uparrow + \rho_\downarrow$). Blue dashed lines indicate z positions of V and Se, while the red dashed line indicates the z position of CoPc. The Pauli pushback effect can be observed by the accumulation of electron charge just above the interface Se atoms.

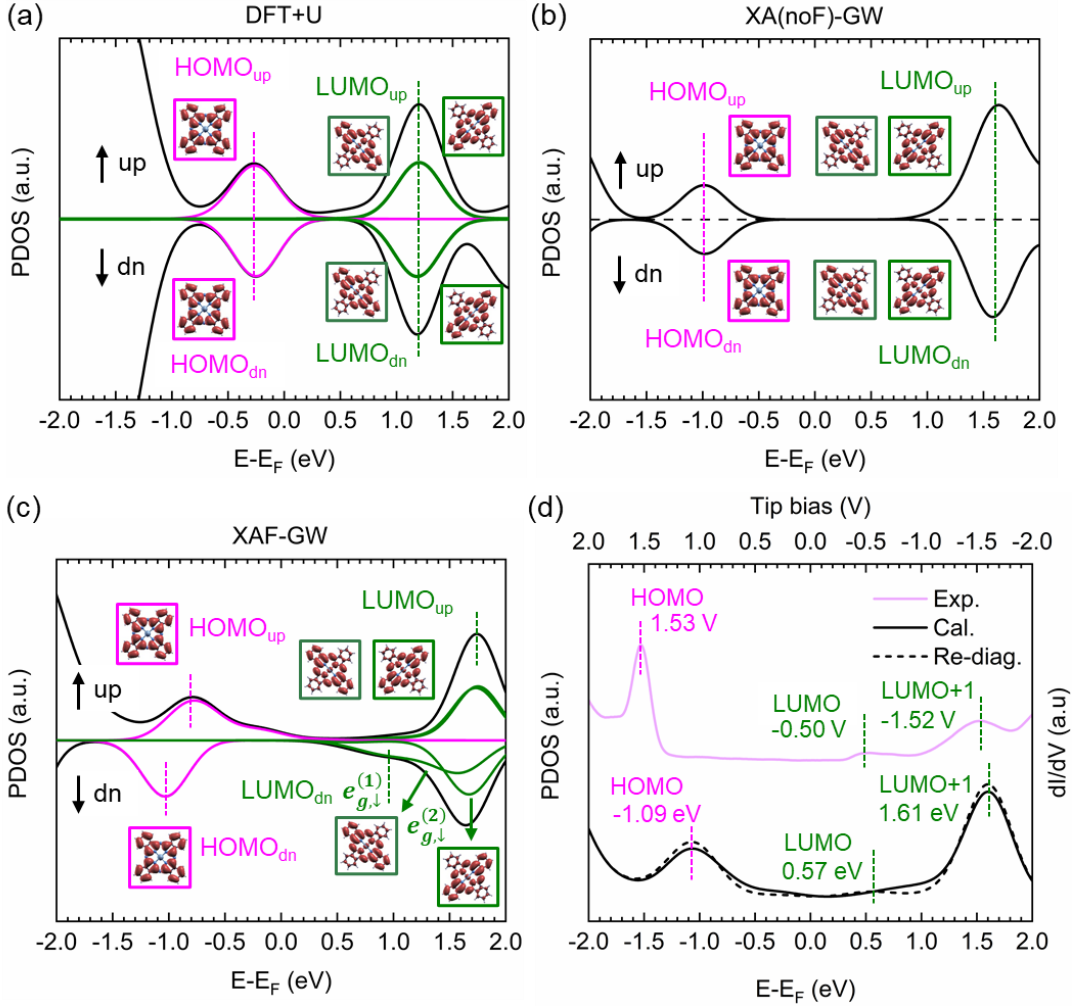

Figure 3. Projected density of states (PDOS) on CoPc in H-site CoPc/VSe$_2$. Spin-polarized PDOS on CoPc using (a) DFT + U, (b) XA(noF)-GW, (c) XAF-GW. (d) Comparison between scanning tunneling spectra (STS)[28] and the total (spin up + spin down) projected density of states (PDOS) for CoPc in H-site CoPc/VSe$_2$ obtained with XAF-GW approach, after shifting the calculated Fermi level of 0.1eV to account for the graphite substrate in the experiment. Dashed: PDOS after rediagonalization to take into account off-diagonal matrix elements in the self-energy. The STS data corresponds to data for 'regular' molecules in Ref. 28. In (d), the MOs are labeled without reference to spin, so that LUMO+1 refers to the LUMO$_{up}$ in (c). PDOS obtained by projecting CoPc/VSe$_2$ wavefunctions onto frontier CoPc wavefunctions ($a_{1u}$ and $e_g$) are indicated by magenta and green curves, respectively, in (a) and (c). The corresponding gas phase molecular orbitals are shown in magenta and green boxes in (a)-(c).

It is known that the ELA involving TMPc molecules is particularly difficult to predict with first principles calculations - DFT calculations with standard exchange-correlation functionals result in wrong ordering of the molecular orbitals (MOs) in TMPc molecules.[32-34] Benchmark calculations on gas phase CoPc molecules have been performed using many-body perturbation theory within the GW approximation, starting from DFT Hamiltonians with a hybrid exchange-correlation functional.[35] This procedure resulted in good quantitative agreement with photoelectron spectroscopy experiments.[35] In this work, we perform GW calculations starting from a DFT Hamiltonian with a Hubbard U correction (see Methods); this procedure results in gas phase MO levels that are in very good agreement with those in Ref. 35 (Fig. S1), providing a computationally more tractable but equally predictive alternative to compute the ELA for frontier orbitals in CoPc/VSe$_2$. To overcome the computational bottleneck in computing the non-interacting polarizability matrix for the interface, we further generalize our recently developed XAF-GW approach[12] to treat interfaces with arbitrary supercells (see Methods). It has been shown both analytically and numerically that this approach is valid even in the presence of interface hybridization.[12]

The DFT+U projected density of states (PDOS) onto CoPc in CoPc/VSe$_2$ is shown in Fig. 3a. Similar to the gas phase, the HOMO is spin-degenerate, while the spin splitting in the LUMO levels is less than 0.05 eV (Table S1). In contrast, the XAF-GW PDOS (Fig. 3c) reveals clearly the presence of spin-dependent tunneling barriers in the scanning probe setup for CoPc/VSe$_2$, indicating that many-body self-energy effects give rise to spin-dependent tunneling barriers. For example, HOMO$_{up}$ (spin up HOMO) is 0.25 eV closer to the Fermi level, E$_F$, than HOMO$_{dn}$ (spin down HOMO), corresponding to a smaller tunneling barrier for spin down holes (using the convention[36] that the spin of the hole is opposite to that of the electron originally occupying the state). This result can be understood from the spin-dependent hybridization between the spin-degenerate gas-phase $a_{1u}$ HOMO and the VSe$_2$ spin up and spin down states, which affects the extent of localization of the hybridized state. Due to the $\frac{1}{r}$ nature of the Coulomb interaction, the self-energy correction (which shifts levels away from E$_F$) is larger for more localized wavefunctions.[37] From the spin-polarized DFT+U PDOS on VSe$_2$ in the CoPc/VSe$_2$ heterostructure (Fig. 1d), we see that the spin up VSe$_2$ PDOS is larger than the spin down PDOS close to E$_F$. As a result, the HOMO$_{dn}$ state is more localized on the CoPc molecule than HOMO$_{up}$ (Fig. 1e), resulting in a larger self-energy correction for the HOMO$_{dn}$ state. Similarly, the $e_{g,\downarrow}^{(1)}$ LUMO state (Fig. 3c) has a smaller self-energy correction than the $e_{g,\downarrow}^{(2)}$ and LUMO$_{up}$ states, because it is less localized on the molecule (Fig. S2). This reasoning is further confirmed from the XA(noF)-GW calculation,[12] where the CoPc monolayer wave functions are used, instead of the heterostructure wavefunctions, to compute the quasiparticle levels. This approach captures the effects of inter-molecular screening and substrate-induced screening on the quasiparticle levels, but not the effect of interface hybridization (see Methods, Fig. 4). We see that the XA(noF)-GW gives spin-degenerate HOMO and LUMO just like DFT+U (Fig. 3b), thus further confirming that it is the interface hybridization that results in a spin-dependence of the self-energies. Our calculations reveal clearly the important role of many-body effects and interface hybridization in determining the spin-dependent tunneling barriers of this mixed-dimensional magnetic heterostructure.

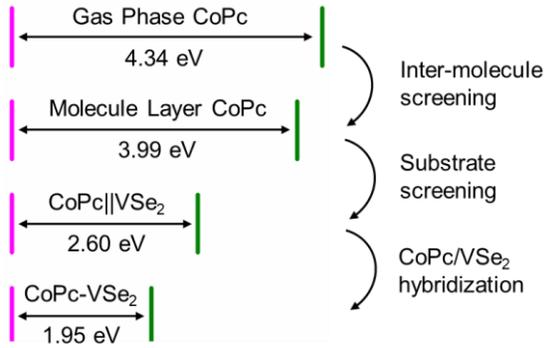

Figure 4. Schematic figure for the evolution of CoPc HOMO-LUMO gap. The pink and green lines stand for HOMO and LUMO, where HOMO position is aligned. CoPc∥VSe$_2$ represents an artificial structure in XA(noF)-GW calculation without hybridization between CoPc and VSe$_2$. CoPc-VSe$_2$ represents the physical system with hybridization, with the HOMO-LUMO gap obtained using XAF-GW. Gas phase and monolayer CoPc gaps are obtained with GW calculations.

Summing the spin up and spin down PDOS on CoPc to compare with the experimental STS data, we obtain a HOMO-LUMO gap of 1.95 eV in CoPc/VSe$_2$, in agreement with the experimental value of 2.03 eV[28] (Fig. 3d). In contrast, the GW HOMO-LUMO gap in gas phase CoPc is much larger (4.34 eV, Fig. S1). Fig. 4 gives a schematic illustration of the different effects that result in a reduction in the HOMO-LUMO gap for CoPc when placed on VSe$_2$. Comparison with the GW HOMO-LUMO gap in monolayer CoPc (3.99 eV) and the XA(noF)-GW HOMO-LUMO gap in CoPc/VSe$_2$ (2.60 eV) shows that both the surrounding molecules and the VSe$_2$ substrate participate in screening quasiparticle excitations in the CoPc molecule, reducing the HOMO-LUMO gap (by 1.74 eV). At the same time, reduced self-energy corrections stemming from interface hybridization effects, which reduce the localization of the wavefunction on the molecule, further reduce the HOMO-LUMO gap by 0.65 eV (XA(noF)-GW versus XAF-GW, Fig. 4). The LUMO for XAF-GW is defined here as the shoulder which arises from reduced self-energy corrections for the interface-hybridized $e_{g,\downarrow}^{(1)}$ LUMO state. This feature is consistent with the shoulder in the experimental spectra (Fig. 3d), providing further evidence of interface hybridization effects on the PDOS. To compare the predicted ELA with experiment, we shift $E_F$ by 0.1 eV to take into account the effect of the graphite substrate in the experiment on the work function (Fig. 3d). The effect of off-diagonal matrix elements in the self-energy are also taken into consideration, resulting in a slightly more prominent shoulder due to the $e_{g,\downarrow}^{(1)}$ LUMO state, in better agreement with experiment. We note that these off-diagonal matrix elements do not change our conclusions on spin-dependent tunneling barriers (see SI, Fig. S4).

In summary, we have shown that the interplay between many-body effects and interface hybridization play an important role in determining the HOMO-LUMO gap of CoPc on $VSe_2$, as well as the experimentally observed shoulder in the STS data corresponding to unoccupied MO states.[28] Simple concepts of screening due to the substrate and the surrounding molecules, which have been adopted previously for physisorbed molecules on substrates,[10-11] are insufficient to explain the experimental[28] observations. These same many-body effects in interface-hybridized states also result in a spin dependence of the tunneling barriers, which are absent in the mean-field predictions. In contrast to spin-quenching observed for CoPc on Pb(111)[3-4] and Au(111)[5] substrates, the magnetic moment on CoPc is slightly enhanced on $VSe_2$, with the Co and V spins antiparallel to one another. Taken together, our results for CoPc/$VSe_2$ indicate that TMPc molecules on 2D layered magnets are a promising platform for the realization of molecular spintronics and quantum computing applications. The importance of interface hybridization in determining the spin-dependent ELA also suggests that spin-dependent tunneling barriers can be controlled by tuning the interface hybridization, using gate voltages or chemical means.[5]

## COMPUTATIONAL METHODS

Following Ref. 38, the $GW$ quasiparticle energy ($E^{QP}$) is defined as:

$$E^{QP} = \langle\psi|H^{DFT+U} + \Sigma - V_{xc} - U|\psi\rangle = E^{DFT+U} + \langle\psi|\Sigma - V_{xc}|\psi\rangle - \langle\psi|U|\psi\rangle,$$

where $\psi$ represents the DFT+U Kohn-Sham eigenstates, $H^{DFT+U}\psi = E^{DFT+U}\psi$, $V_{xc}$ is the exchange correlation potential in the DFT Hamiltonian and $\langle\psi|U|\psi\rangle = \langle\psi|H^{DFT+U} - H^{DFT}|\psi\rangle = E^{DFT+U} - \langle\psi|H^{DFT}|\psi\rangle$. $E^{DFT+U}$, $\psi$ and $\langle\psi|U|\psi\rangle$ are obtained from the DFT+U starting point. The self-energy correction, $\langle\psi|\Sigma - V_{xc}|\psi\rangle$, is calculated within the Berkeley$GW$ package[16] using the one-shot $G_0W_0$ approximation, together with the generalized plasmon pole model,[16] and a slab Coulomb truncation scheme.[39] The static limit of the dielectric function is computed by the Random Phase Approximation (RPA), $\epsilon^{RPA} = 1 - v\chi^0$, where $\chi^0$ is the independent particle polarizability. A $G_1W_1$ calculation of the PDOS is provided in Fig. S5. For the CoPc/VSe$_2$ heterostructure, the Brillouin Zone (BZ) is sampled with a $2 \times 2 \times 1$ k-mesh. 2000 bands are used in the evaluation of the Green's function and a 10 Ry cutoff for $\chi^0$ is used. The gas phase CoPc gap is converged with an equivalent kinetic energy cutoff, and with a $10\,Ry$ cutoff for $\chi^0$. Reducing the number of bands to 1800 for the CoPc/VSe$_2$ heterostructure does not change the PDOS. The effects of off-diagonal GW Hamiltonian matrix elements are taken into account as described in the SI.

For CoPc/VSe$_2$, we apply the XAF-GW method[12] where the only approximation is:

$$\chi^{0,HS}_{\vec{G}\vec{G}'}(\vec{q}) = \chi^{0,CoPc}_{\vec{G}\vec{G}'}(\vec{q}) + \chi^{0,VSe2}_{\vec{G}\vec{G}'}(\vec{q})$$

This approximation is valid in the presence of interface hybridization, in the absence of strong covalent bonds and where the MOs do not cross the Fermi level,[12] conditions that hold for CoPc/VSe$_2$.

The XAF-GW approach consists of three steps (Fig. 1f).

In the first step, $\chi^0$ is computed for the smallest possible unit cell of each component of the interface, and then expanded to that for the supercell. In this work, the VSe$_2$ supercell is given by: $\vec{a}_1 = 6\vec{A}_1 + 3\vec{A}_2$, $\vec{a}_2 = \vec{A}_1 + 5\vec{A}_2$, where $\vec{A}_1$ and $\vec{A}_2$ are the lattice vectors of the VSe$_2$ unit cell (Fig. 1b), corresponding in reciprocal space (Fig. 1c) to: $\vec{B}_1 = 6\vec{b}_1 + \vec{b}_2$, $\vec{B}_2 = 3\vec{b}_1 + 5\vec{b}_2$, where $\vec{B}_1$ and $\vec{B}_2$ are the reciprocal space lattice vectors of the VSe$_2$ unit cell. Using:

$$\chi^0(\vec{r};\vec{r}') = \sum_{\vec{q}\vec{G}\vec{G}'} e^{i(\vec{q}+\vec{G})\vec{r}} \chi^0_{\vec{G}\vec{G}'}(\vec{q}) e^{-i(\vec{q}+\vec{G}')\vec{r}'}$$

$$= \sum_{\vec{q}\vec{G}_u\vec{G}'_u} \sum_{\vec{G}_I} e^{i(\vec{q}+\vec{G}_I+\vec{G}_u)\vec{r}} \chi^0_{\vec{G}_u\vec{G}'_u}(\vec{q}+\vec{G}_I) e^{-i(\vec{q}+\vec{G}_I+\vec{G}'_u)\vec{r}'}$$

one can find that each supercell chi matrix element $\chi^0_{\vec{G}\vec{G}'}(\vec{q})$ corresponds to a set of unit cell chi matrix elements $\chi^0_{\vec{G}_u\vec{G}'_u}(\vec{q}+\vec{G}_I)$, where $\vec{G}_I$ runs over the set of all the supercell reciprocal space vectors such that $(\vec{q}+\vec{G}_I)$ falls in the first BZ for the unit cell (see Fig. 1c for the dots inside the blue parallelogram formed by $\vec{B}_1, \vec{B}_2$). This is a generalized version of the approach presented in Ref. 12.

In the second step, chi matrices for different components are added together.

In the third step, we use the heterostructure wave function to evaluate the self-energy correction which takes into account the hybridisation between CoPc and $VSe_2$, while for XA(noF)-GW, we use the CoPc monolayer wave function, which artificially neglects the interlayer hybridisation.

## ACKNOWLEDGEMENTS

We acknowledge funding from Grant MOE2016-T2-2-132 from the Ministry of Education, Singapore and support from the Singapore National Research Foundation, Prime Minister's Office, under its medium-sized centre program. We thank A. T. S. Wee and L. Zhang for sending us their experimental data. Computations were performed on the NUS Graphene Research Centre cluster and National Supercomputing Centre Singapore (NSCC). We thank M. D. Costa for systems support.

## AUTHOR CONTRIBUTIONS

Runrun Xu and Fengyuan Xuan contributed equally to this work.

**Supporting Information Available:** Further details of methods, GW results for gas phase CoPc, Wave functions, PDOS and Energy levels, $G_1W_1$ PDOS for CoPc/$VSe_2$ This material is available free of charge *via* the Internet at http://pubs.acs.org.

# Supporting information for "Spin-dependent Tunneling Barriers in CoPc/VSe$_2$ from Many-Body Interactions"

*Runrun Xu[1,†], Fengyuan Xuan[2,†], and Su Ying Quek[1,2]*  *

[1]Department of Physics, National University of Singapore, 2 Science Drive 3, 117542, Singapore

[2]Centre for Advanced 2D Materials, National University of Singapore, Block S14, Level 6, 6 Science Drive 2, 117546, Singapore

* To whom correspondence should be addressed: phyqsy@nus.edu.sg

†: These authors contributed equally to this work.

**Contents:**

**1. Geometry optimization**

**2. Hubbard U and magnetic moments**

**3. Mean-field DFT starting point for GW**

**4. GW results for gas phase CoPc**

**5. Gamma-point wave functions of CoPc/VSe$_2$**

**6. PDOS on CoPc in CoPc/VSe$_2$ and corresponding molecule orbitals**

**7. Table showing different molecule orbital levels**

**8. Effect from off-diagonal GW Hamiltonian matrix elements**

**9. G$_1$W$_1$ PDOS for CoPc in CoPc/VSe$_2$**

## 1. Geometry optimization

Geometry optimization is carried out using the Vienna *ab-initio* simulation package (VASP),[1] where the electron-core interactions are described by the projector augmented wave (PAW)[2] method. The exchange correlation energy is treated using Perdew-Burke-Ernzerhof generalized gradient approximation (PBE-GGA).[3] Grimme DFT-D2 corrections[4] are applied to account for *vdW* interactions in the heterostructure. The energy cut-off for plane waves is set to be $400\ eV$ and the criterion for the total energy is $1.0 \times 10^{-6}\ eV$. All the ionic positions are optimized until the forces acting on ions become less than $0.01$ and $0.02\ eV/Å$ for the single molecule and the heterostructure, respectively.

Our optimized gas phase CoPc molecular geometry is in good agreement with experimental measurements[5] and previous theoretical calculations,[6] with a Co-N bond length (the bond connecting the central Co atom and the nearest N element) of 1.92 Å, (experiment: 1.91 Å).[5] Our monolayer T-phase $VSe_2$ has an optimized in-plane lattice constant of 3.32 Å (experiment: $3.31 \pm 0.05$ Å).[7]

In order to simulate the $CoPc/VSe_2$ heterostructure, we adsorb each CoPc molecule on a monolayer $VSe_2$. The vacuum length is at least 15 Å and dipole corrections[2,8] are applied, making sure that there are no artificial interactions between neighbouring slabs due to the periodic boundary conditions. Three adsorption sites are considered here: "V-site" with Co on top of one V atom in underlying $VSe_2$; "Se-site" with Co on top of one Se atom in underlying $VSe_2$ and "H-site" with Co above a hollow site. After a full relaxation of the atomic positions, the interface separations (defined as the distance between central Co atom and the interface Se layer) are optimized to be ~ 3.27, 3.16 and 3.25 Å for V-, Se- and H- site, respectively.

## 2. Hubbard U and magnetic moments

With the optimized geometries, magnetic moments are obtained in VASP by applying a Hubbard $U$ parameter to the DFT Hamiltonian[9] to partially account for the strongly correlated nature of the localized $3d$ electrons on Co and V. The adsorption energies reported in the main text were obtained using the PBE exchange-correlation functional[3] with a Hubbard U correction to the Hamiltonian,[10] taking $U = 3.00\ eV$ for V $3d$ orbitals[11-12] and $U = 5.00\ eV$ for the Co $3d$ orbitals. The magnetic moment in the pristine VSe$_2$ monolayer was 1.28 $\mu_B$ per f.u. and 1.60 $\mu_B$ per V atom, similar to previous reports,[11-13] and that in the gas phase CoPc molecule was 0.95 $\mu_B$ per molecule and 1.04 $\mu_B$ per Co atom. $U = 5.00\ eV$ was chosen for Co because Ref. [6] had found that $U = 4 - 6\ eV$ provided the best fits to valence-band photoelectron spectra for CoPc. The H-site is found to be energetically most favourable for all values of $U$ that we have used in this work.

## 3. Mean-field DFT starting point for GW

The DFT wavefunctions for GW calculations and the PDOS were obtained using the Quantum ESPRESSO package,[14] where optimized norm-conserving pseudopotentials[15] are used with 17 and 13 valence electrons for Co and V, respectively. The plane-wave cutoff is set to 60 $Ry$ [15] and Perdew-Burke-Ernzerhof generalized gradient approximation[3] (PBE-GGA) is adopted for the exchange-correlation energy functional. A Hubbard U is used to take into account the strongly correlated nature of the localized $3d$ electrons on Co and V as described in the next section.

## 4. GW results for gas phase CoPc

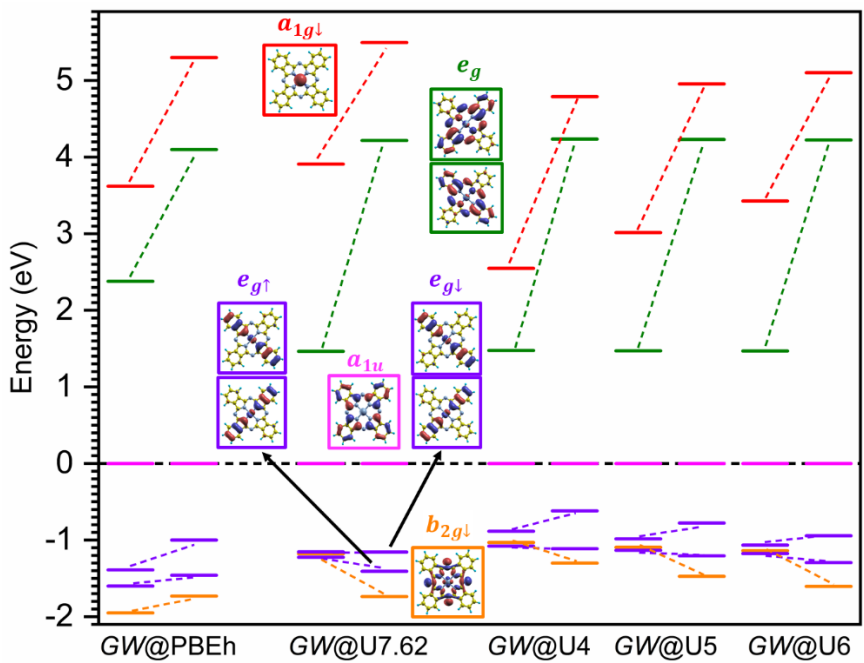

Figure S1. DFT mean field (left) and GW (right) levels for gas phase CoPc using mean-field starting points, the hybrid functional PBEh (*GW*@PBEh) and DFT+U (*GW*@U) wavefunctions with $U = 4, 5, 6$ and $7.62\ eV$ (U4, U5, U6 and U7.62, respectively) for Co. The levels are aligned according to the HOMO (magenta). The data for *GW*@PBEh is taken from Ref. [16]. The XAF-GW calculations for CoPc/VSe$_2$ use $U$ = 7.62 eV for Co. Color codes represent the different MOs, and dashed lines are a visual guide to show the effect of the GW self-energy correction after the HOMO is aligned.

The $a_{1u}$ HOMO and $e_g$ LUMO are primarily localized on the arms of the molecule, away from the Co center, and the HOMO-LUMO gaps all closely match that from GW@PBEh[16] for different values of *U*. The energies of other molecular orbitals (MOs) with more weight on Co are more sensitive to the choice of *U*. The GW@U7.62 MO energies are in very good agreement with those of GW@PBEh and with experiment,[16-17] and noting that $U = 7.62$ eV is the self-consistent value of $U$ computed using linear response[18-19] for CoPc/VSe$_2$, the XAF-GW calculations for CoPc/VSe$_2$ use $U = 7.62$ eV for Co (and $U = 3.00\ eV$ for V).[11-12]

## 5. Gamma-point wave functions of CoPc/VSe$_2$

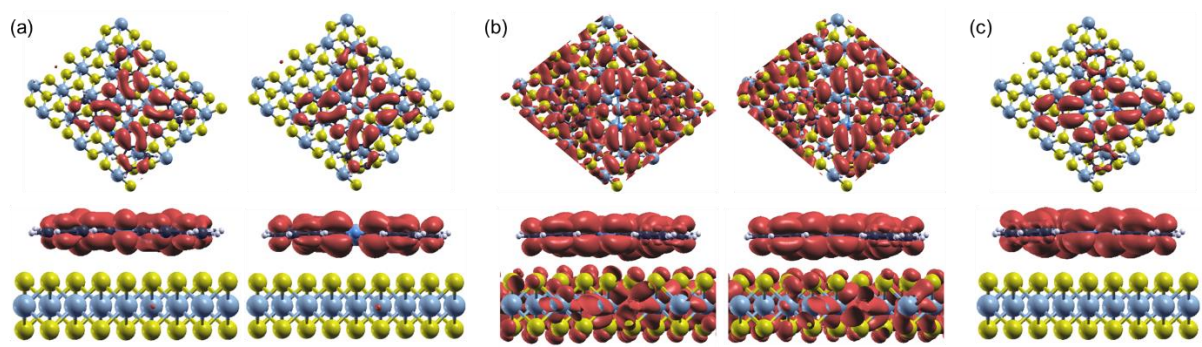

Figure. S2. Wave functions of CoPc/VSe$_2$ heterostructure at the Γ-point of Brillouin zone: (a) nearly degenerate LUMO$_{up}$ states, (b) $e_{g,\downarrow}^{(1)}$ and (c) $e_{g,\downarrow}^{(2)}$, as labeled in Figure 3c in the main text. The isosurface values are taken to be 1% of the maximum value. Note that in (b), the gas phase molecule orbital projects onto two heterostructure wave functions at the Γ-point.

## 6. PDOS on CoPc in CoPc/VSe$_2$ and corresponding molecule orbitals

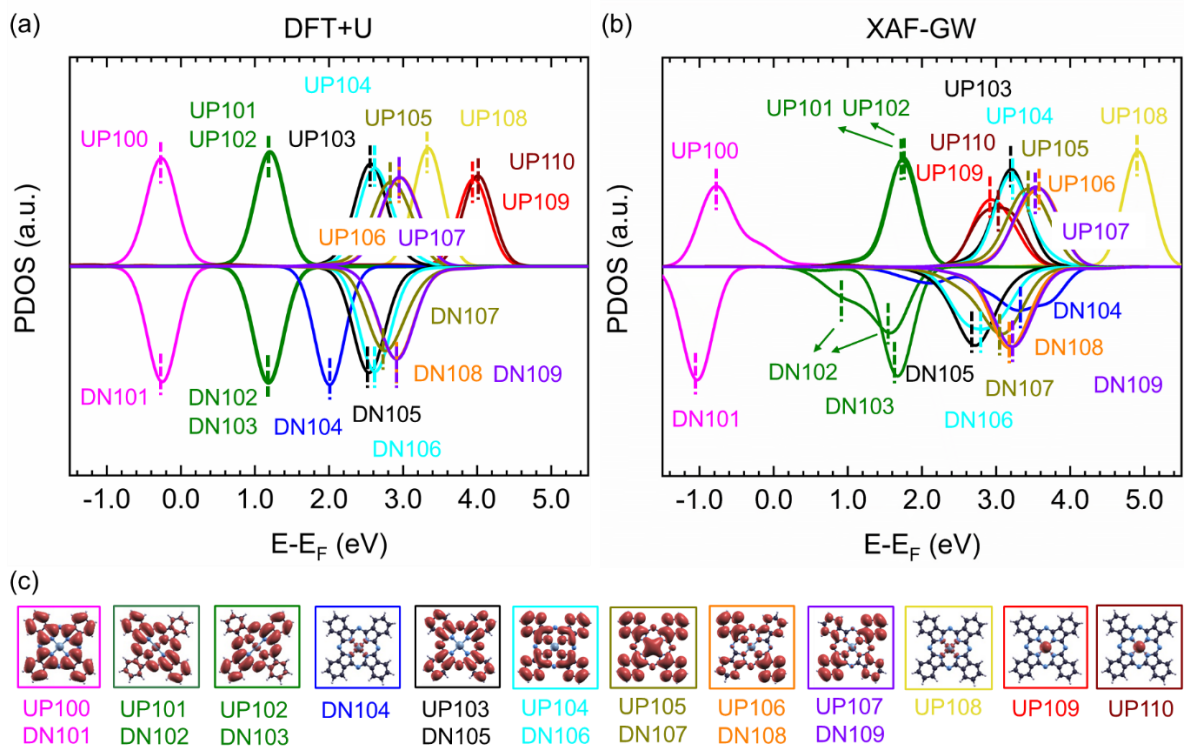

Figure S3. Projected density of states (PDOS) onto different individual molecule orbitals (MO) of the CoPc monolayer, using (a) DFT+U and (b) XAF-GW approach, corresponding to the PDOS in Fig. 3(a) and Fig. 3(c) of the main text, respectively. The wave functions of the corresponding MO in the CoPc monolayer are shown in (c). 'UP' and 'DN' refer to spin up or spin down, and the number indicates the index of the gas-phase MO. HOMO$_{up}$ corresponds to UP100 and HOMO$_{dn}$ corresponds to DN101.

From Fig. S3 and Table S1 below, we see that the GW ordering in CoPc/VSe$_2$ can differ from the DFT+U one, and also that the GW ordering in CoPc/VSe$_2$ can differ from that in gas phase CoPc.

## 7. Table showing different molecular orbital levels

Table S1. Selected MO levels in $eV$ for gas phase CoPc and for CoPc/VSe$_2$. 'UP' and 'DN' refer to spin up or spin down, and the number indicates the index of the gas-phase MO. **HOMO$_{up}$ corresponds to UP100 and HOMO$_{dn}$ corresponds to DN101**. Mean field DFT+U calculations were performed using $U = 7.62\ eV$ for Co and $U = 3.00\ eV$ for V. $\Delta E_{spin}$, defined as $\Delta E_{spin} = E_{up} - E_{dn}$, refers to the spin splitting for the same MO. The colors in the table correspond to the colors in Fig. S3. The two values for DN102 in H-site CoPc/VSe$_2$ refer to the two peaks of this MO as shown in Fig. 3(c). Energies are aligned with respect to the Fermi level (or the HOMO in gas phase). We observe that the GW ordering of levels in in CoPc/VSe$_2$ can be different from that in gas phase CoPc (eg. for DN104 and UP110).

|        | Gas Phase CoPc (GW) |                    | CoPc/VSe$_2$ (DFT) |                    | CoPc/VSe$_2$ (XAF-GW) |                    |
|--------|---------------------|--------------------|--------------------|--------------------|-----------------------|--------------------|
|        | E                   | $\Delta E_{spin}$  | E                  | $\Delta E_{spin}$  | E                     | $\Delta E_{spin}$  |
| UP100  | 0.00                | -0.01              | -0.26              | 0.00               | -0.77                 | 0.27               |
| DN101  | 0.01                |                    | -0.26              |                    | -1.04                 |                    |
| UP101  | 4.38                | 0.04               | 1.20/1.21          | 0.04/0.01          | 1.74/1.76             | 0.77/0.20          |
| DN102  | 4.34                |                    | 1.16/1.20          |                    | 0.97/1.56             |                    |
| UP102  | 4.38                | 0.04               | 1.20/1.21          | 0.02/0.03          | 1.74/1.76             | 0.07/0.09          |
| DN103  | 4.34                |                    | 1.18               |                    | 1.67                  |                    |
| DN104  | 6.53                | ---                | 2.01               | ---                | 3.29                  | ---                |
| UP108  | 7.74                | ---                | 3.34               | ---                | 4.91                  | ---                |
| UP109  | 5.01                | ---                | 3.86               | ---                | 2.95                  | ---                |
| UP110  | 5.53                | ---                | 4.04               | ---                | 3.03                  | ---                |

## 8. Effects from off-diagonal GW Hamiltonian matrix elements

In our XAF-GW calculation, we first perform a DFT calculation with the Hubbard U potential:

$$H^{DFT+U}\psi_{nk} = E_{nk}^{DFT+U}\psi_{nk}$$

The quasiparticle (QP) energies are then calculated using

$$E_{nk}^{QP} = \langle\psi_{nk}|H^{GW}|\psi_{nk}\rangle = \langle\psi_{nk}|H^{DFT+U} + \Sigma - V_{xc} - U|\psi_{nk}\rangle$$

Details of this calculation are provided in Methods (main text). However, the above expression assumes that the GW Hamiltonian is still diagonal in the basis of DFT wavefunctions, that is, the off-diagonal elements:

$$\langle\psi_{nk}|H^{GW}|\psi_{mk}\rangle = \langle\psi_{nk}|\Sigma - V_{xc} - U|\psi_{mk}\rangle \approx 0$$

where $n \neq m$. In other words, the expression assumes that the DFT wave function is a good approximation to the GW wave function, which is generally a very good approximation.[20] However, we observe that when two heterostructure (HS) states, $\psi_{nk}$ and $\psi_{mk}$ both have a large projection on the same molecule state $\psi_k^{CoPc}$, the off-diagonal GW Hamiltonian matrix element $\langle\psi_{nk}|H^{GW}|\psi_{mk}\rangle$ is not negligible. Therefore, we need to re-diagonalize the sub-space of these two HS states:

$$\begin{bmatrix}\langle\psi_{nk}|H^{GW}|\psi_{nk}\rangle & \langle\psi_{nk}|H^{GW}|\psi_{mk}\rangle \\ \langle\psi_{mk}|H^{GW}|\psi_{nk}\rangle & \langle\psi_{mk}|H^{GW}|\psi_{mk}\rangle\end{bmatrix}\begin{bmatrix}c_1 \\ c_2\end{bmatrix} = \begin{bmatrix}\tilde{E}_{nk}^{QP} & 0 \\ 0 & \tilde{E}_{mk}^{QP}\end{bmatrix}\begin{bmatrix}c_1 \\ c_2\end{bmatrix}$$

and compare the new QP energies, $\tilde{E}_{nk}^{QP}$, $\tilde{E}_{mk}^{QP}$, with the original ones, $E_{nk}^{QP}$, $E_{mk}^{QP}$.

Fig. S4 below shows the result of this rediagonalization procedure for all frontier molecular states (dashed lines. We can see that the off-diagonal elements slightly modify the peak positions and shapes in the PDOS, especially for HOMO$_{up}$ and LUMO$_{dn}$, for which the off-diagonal elements are non-negligible.

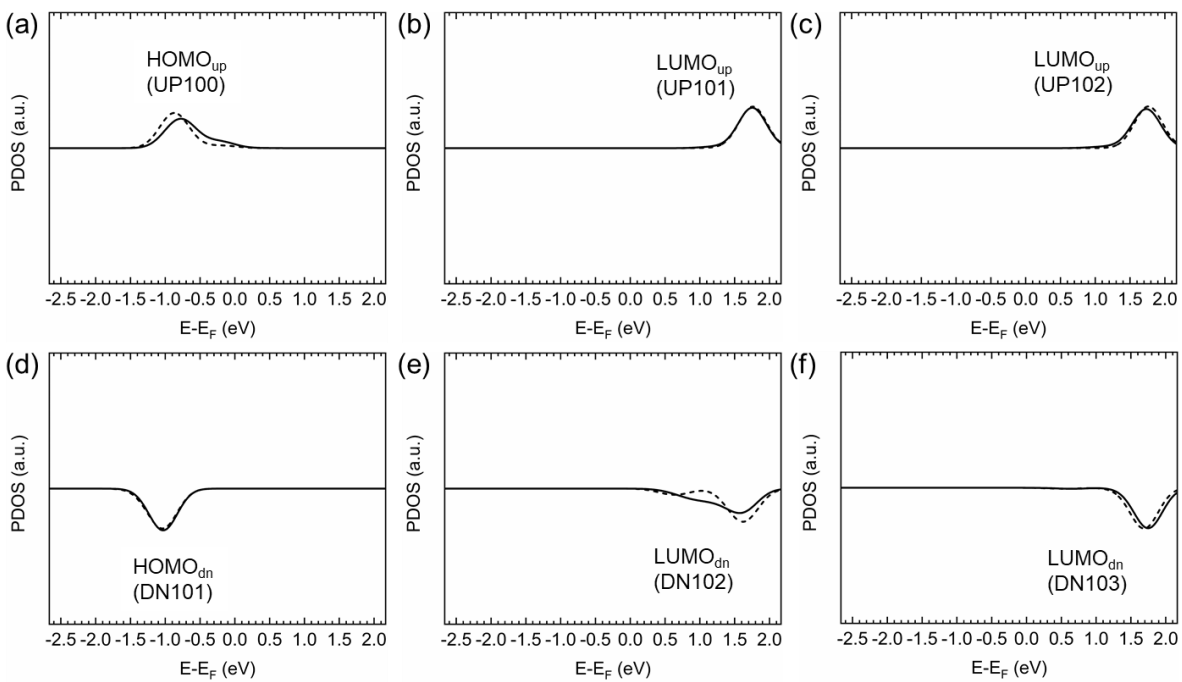

Figure S4. PDOS on frontier molecular orbitals without (solid) and with (dashed) the rediagonalization process described above. Numerical indices refer to the band index of the gas phase molecule.

## 9. $G_1W_1$ PDOS for CoPc in CoPc/VSe$_2$

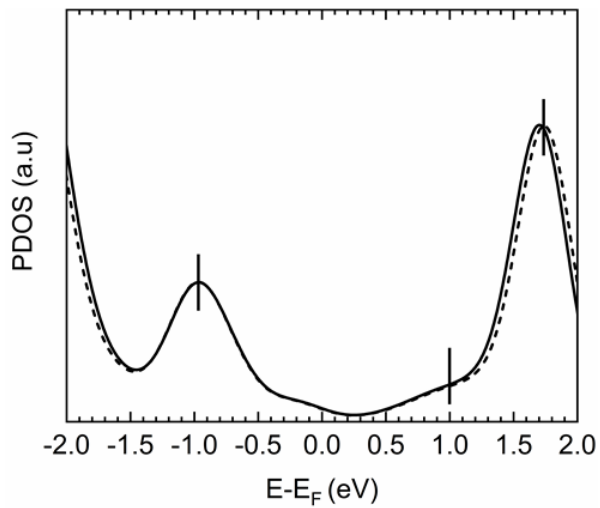

Figure S5. $G_0W_0$ (solid) and $G_1W_1$ (dashed) PDOS for CoPc in CoPc/VSe$_2$.